\documentclass[reprint,aps, superscriptaddress,nofootinbib]{revtex4-1}

\usepackage{graphicx}
\usepackage{dcolumn}
\usepackage{bm}

\usepackage{amsmath}
\usepackage{amssymb}
\usepackage{amsfonts}
\usepackage{amsthm}
\usepackage{color}
\usepackage{subfig}

\begin{document}

\captionsetup{justification=raggedright,singlelinecheck=false}

\title{A Stern--Gerlach separator of chiral enantiomers based on the Casimir-Polder potential}

\author{Fumika Suzuki}
 \email{fumika@chem.ubc.ca}
  \affiliation{Department of Chemistry, University of British Columbia, Vancouver, British Columbia, Canada V6T 1Z1}
 \affiliation{%
Department of Physics and Astronomy, University of British Columbia, Vancouver, British Columbia, Canada V6T 1Z1
}
  \affiliation{Physikalisches Institut, Albert-Ludwigs-Universit\"{a}t Freiburg, Hermann-Herder-Str. 3, 79104 Freiburg, Germany}
\author{Takamasa Momose}
 \email{momose@chem.ubc.ca}
  \affiliation{Department of Chemistry, University of British Columbia, Vancouver, British Columbia, Canada V6T 1Z1}
  \author{S. Y. Buhmann}
 \email{stefan.buhmann@physik.uni-freiburg.de}
  \affiliation{Physikalisches Institut, Albert-Ludwigs-Universit\"{a}t Freiburg, Hermann-Herder-Str. 3, 79104 Freiburg, Germany}
  \affiliation{Freiburg Institute for Advanced Studies, Albert-Ludwigs-Universit\"{a}t Freiburg, Albertstr. 19, 79104 Freiburg, Germany}
\date{\today}

\begin{abstract}
We propose a method to separate enantiomers using parity violation in the Casimir--Polder potential between chiral mirrors and chiral molecules. The proposed setup involves a molecular beam composed of chiral molecules passing through a planar cavity consisting of two chiral mirrors. Enantiomers of opposite handedness are deflected differently due to a chiral dependence of the Casimir--Polder potential resulting in the separation of the enantiomers. Our setup provides an alternative experimental tool for enantiomer separation, as well as shedding light on the fundamental properties of the Casimir-Polder potential.
\end{abstract}

\maketitle


\section{Introduction}

Many molecules are chiral which can exist in left- and right-handed forms (i.e., non-superimposable mirror images). These two forms of a chiral molecule
are known as enantiomers. Distinguishing two types of enantiomers is of great practical importance. For example, in designing pharmaceuticals, it is necessary to choose the right enantiomer to obtain the desired effects since the other enantiomer is less active, inactive, or can even have adverse side effects, including high toxicity. Therefore, developing technology which separates enantiomers can improve pharmacy by introducing medicines composed of only one enantiomer which enhances the desired effects, while eliminating the side effects \cite{saito}.

Common methods for chiral separation include chromatography \cite{ward} and crystalization \cite{wang}. Chiral separations using chromatography now represent a popular, robust routine technique utilized in laboratories.  However, the selection of columns  still remains a matter of trial and error, and it is difficult to find materials that show both high efficiency and high enantio-selectivity. 
Stern--Gerlach type separators for enantiomers have been proposed that are based on inhomogeneous laser fields \cite{bruder,shapiro}. The trajectories of the emerging molecular beams depend on both spin and handedness. In such schemes, the orientation and rotation of the molecules needs to be carefully addressed \cite{hornberger}. 

Here we propose a method for the  separation of chiral enantiomers that is based on the Casimir-Polder (CP) potential. CP forces are effective quantum electrodynamical forces between neutral, unpolarised molecules and macroscopic bodies which arise from the interaction of the objects' charge and current densities with the vacuum electromagnetic field. Originally predicted by Casimir and Polder, they are typically attractive and purely monotonic for ground-state molecules and bodies with a purely electric response \cite{Casimir48}. It was later seen that both the direction and distance dependence of the potential changes when considering excited molecules. Here, the possibility that the molecule emits a real photon which gets reflected off the body and subsequently reabsorbed leads to a potential whose sign depends on the relative phase of emitted and reabsorbed photons \cite{Wylie85,Ducloy}. Depending on the detuning between the radiation emitted by the molecule and the plasmonic resonances of the body, the potential can be attractive or repulsive in the near zone \cite{Failache} while oscillating with distance in the far zone \cite{Blatt}; and the potential may strongly exceed its ground-state counterpart. 
 \begin{figure}
{%
\includegraphics[clip,width=1\columnwidth]{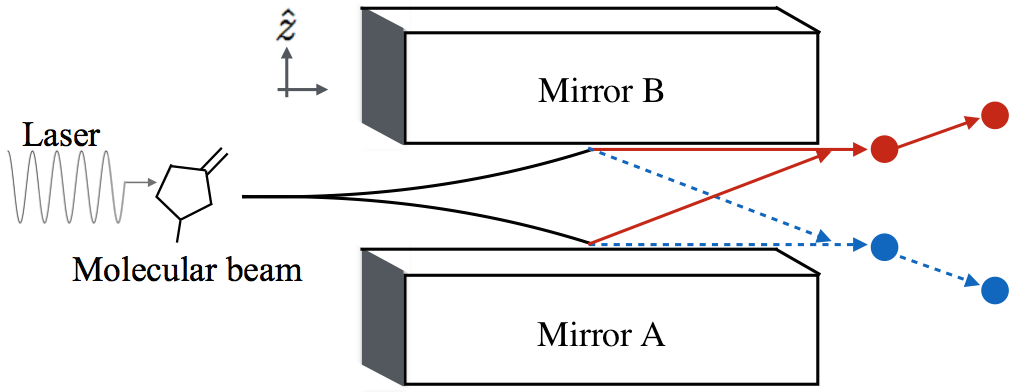}%
}
\caption{(Color online) 
A beam of chiral molecules driven by a laser field passes through a planar cavity consisting of two chiral mirrors. Enantiomers of opposite handedness are deflected differently due to the chiral component of  the CP potential.}
\label{fig1}
\end{figure}

In addition, the sign of the CP force may depend on the electromagnetic response of the interacting objects. Motivated by similar findings for the Casimir force by Boyer \cite{Boyer}, it was found that CP forces due to interactions of electric atoms with magnetic surfaces---or vice versa---are repulsive \cite{Spruch, Kampf, Haakh} in contrast to the attractive electric--electric force. Similar repulsive forces have been predicted for circularly polarised atoms interacting with axionic topological insulators \cite{Fuchs} and---most crucially for this work---for chiral molecules interacting with chiral surfaces. Here the forces are attractive for objects of the same handedness and repulsive for opposite handedness \cite{buh3}.

The chiral component of the CP potential is hence sensitive to the handedness of molecules and can be attractive or repulsive depending on their chirality. This is because the CP potential between a chiral molecule and a chiral mirror depends on the optical rotatory strength of the molecule which is a pseudoscalar that changes sign under a parity inversion \cite{parity}. The potential could hence be used to differentiate enantiomers of opposite handedness. 
Using this property of the CP potential, it would be possible to identify the handedness of the chiral molecule by setting up a Stern-Gerlach type discriminator for enantiomers. Although the electric component of the CP force has been already observed \cite{cp}, here we aim to measure its chiral component which has not yet been  observed. 

The closely related chiral van der Waal force between two molecules has been discussed earlier, where again, enantiomer-discriminatory interactions are predicted when both particles are chiral \cite{sucher, khriplovich1, khriplovich2, salamb}. Since these forces are very weak, a stronger force between a chiral molecule  and an achiral molecule has been proposed which is mediated by a nearby chiral surface \cite{pablo}.

Our proposed
setup uses the molecular beam deflection technique. We let a molecular beam composed of chiral molecules pass through a planar cavity consisting of two chiral mirrors (FIG.~\ref{fig1}). Although the electric component of the CP force between the cavity and the molecules acts on both enantiomers in the same way, the chiral component can be attractive or repulsive depending on the handedness of the molecules  and  can deflect enantiomers of opposite handedness differently and separate them. To exploit the enhanced excited-state CP force, the molecules are subject to a weakly detuned driving laser while passing through the cavity.  The method proposed here is more universal than the above-mentioned conventional schemes, since we only utilize the interaction between a chiral mirror and chiral molecules  induced by photo excitations and emissions.
All molecules have characteristic excitation spectra, and therefore a suitable, resonant light source (either UV or IR) is the only prerequisite for the present method. The same chiral mirror can be used for enantiomer separation of
a wide range of molecules.

The article is organized as follows: In Sec. \ref{sec2}, we study the behavior of the CP potential between the cavity and  the chiral molecule driven by a laser field within the framework of macroscopic quantum electrodynamics. We compare the CP potential experienced by 3-methyl-cyclopentanone (3-MCP) molecules with electric circular dichroism (ECD) and propylene-oxide molecules with vibrational circular dichroism (VCD). We find that the CP potential for propylene-oxide molecules with VCD  can depend on temperature. We also observe the enhancement of the CP potential due to the cavity structure when the cavity is small enough that it contains only a few molecular transition wavelengths. However, experimentally we use a realistic setup where the cavity width is $1\,$mm and such enhancement can not be observed. In Sec. \ref{sec3}, we simulate separation of enantiomers using our setup by studying trajectories of the molecules in the cavity. Since the strength of the CP potential depends on the frequency of photons and the electric and chiral components depend on the electric dipole strength and the optical rotatory strength of the molecules respectively, we use chiral molecules with ECD whose optical rotatory strength is large compared to its electric dipole moment in the simulation. In this article, we use 3-methyl-cyclopentanone as an example for a quantitative analysis. Finally we conclude that our proposed setup can be used for separating enantiomers of opposite handedness and detection of the chiral dependence of the CP potential within reach of current technology (Sec. \ref{sec4}).

\begin{figure}
{%
\includegraphics[clip,width=1\columnwidth]{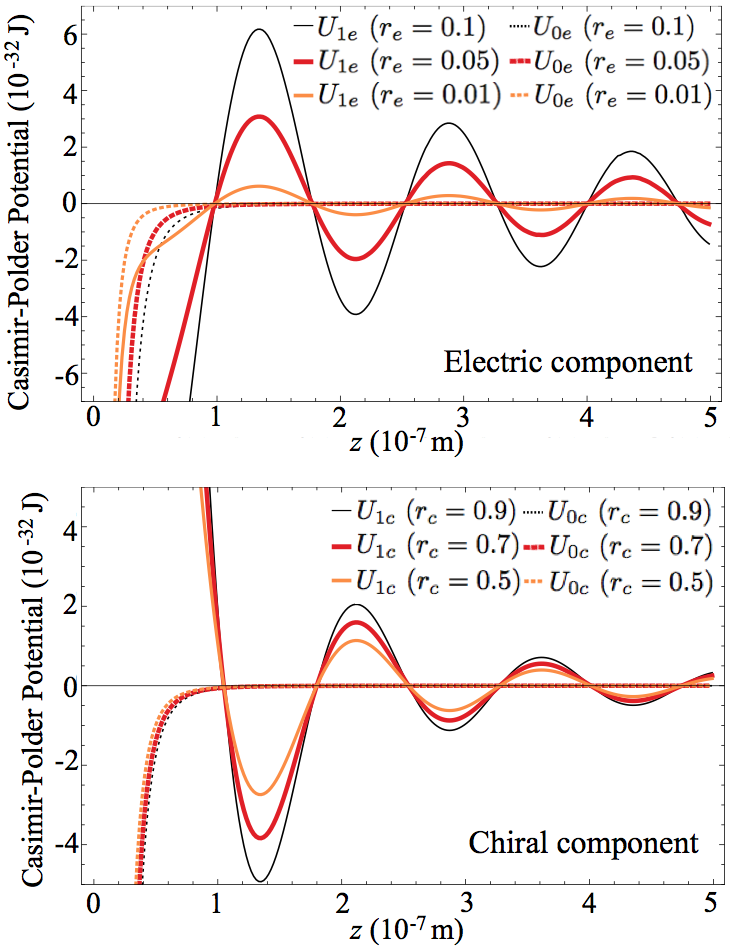}%
}
\caption{(Color online)  The electric and chiral component (experienced by the  3-MCP molecule with the positive $R_{01}$) of  the CP potential with different reflection coefficients: $r_{e}=0.01$, $0.05$ and $0.1$, $r_{c}=0.5$, $0.7$ and $0.9$. The solid lines: the excited-state potential, the dotted lines: the ground-state potential.}
\label{fig2}
\end{figure}

\section{Behavior of the Casimir-Polder Potential}\label{sec2}

\begin{figure}
{%
\includegraphics[clip,width=1\columnwidth]{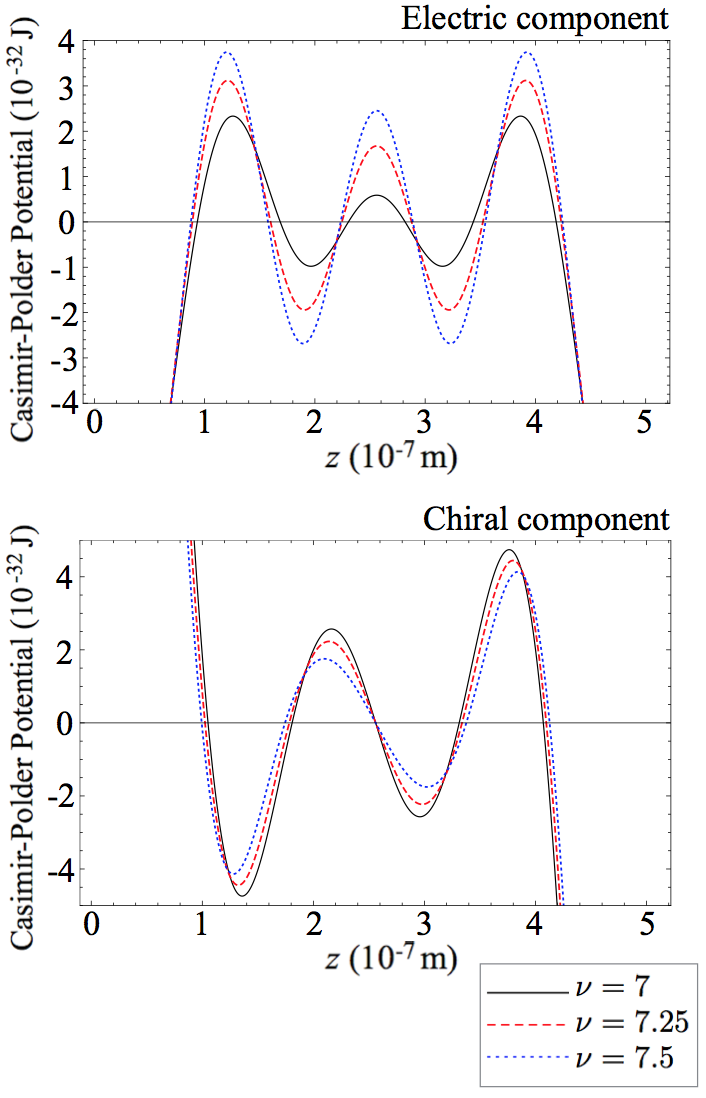}%
}
\caption{(Color online)  The enhancement of the chiral component occurs when $a=\nu  \lambda_{10}/4$ and $\nu$ is odd natural number if the two mirrors have opposite chirality and the cavity width 
contains only a few molecular transition wavelengths. $r_{e}=0.05$, $r_{c}=0.8$.}
\label{fig7}
\end{figure}

Within the framework of macroscopic quantum electrodynamics \cite{buh, buh2}, 
the CP potential $U_{CP}$ between the cavity and the chiral molecule (approximated as two-level system) driven by a laser field is given by \cite{se,se2}
\begin{eqnarray}\label{CPpotential}
U_{CP}=p_0 (t) U_0 +p_1 (t) U_1
\end{eqnarray}
where the
populations
of the ground state $p_0 (t)$ and that of the excited state $p_1 (t)$ are written as
\begin{eqnarray}
p_0 (t)&=& \frac{\Omega^2}{\Delta^2 +\Omega^2}\cos^2 \left(\frac12 \sqrt{\Delta^2 +\Omega^2} \mbox{ }t\right)+\frac{\Delta^2}{\Delta^2+\Omega^2},\nonumber\\
p_1 (t)&=& \frac{\Omega^2}{\Delta^2+\Omega^2}\sin^2 \left(\frac12 \sqrt{\Delta^2 +\Omega^2} \mbox{ }t\right).
\end{eqnarray}
Note that the coherence of the two states gives rise to an additional term which oscillates with optical frequencies and is hence unobservable. 
Here $\Omega$ is the Rabi frequency and the detuning $\Delta=\omega_{L}-\omega_{10}$. $\omega_{10}$ is the molecular transition frequency and $\omega_{L}$ is the driven frequency of the laser. We choose $\omega_{L} \approx \omega_{10}$ to ensure a large excited-state population. 
The potential for the molecule in the ground state is $U_{0}=U_{0e}+U_{0c}$, and that for the molecule in the excited state is $U_{1}=U_{1e}+U_{1c}$, where the electric components of the potential are given by 
\begin{eqnarray}\label{electric}
U_{0e}&=&\frac{\hbar \mu_0}{2\pi}\int^{\infty}_0 d\xi \xi^2 \alpha (i\xi) \mbox{tr}\mathbf{G} (\mathbf{r},\mathbf{r}, i\xi),\nonumber\\
U_{1e}&=&-\frac{\hbar \mu_0}{2\pi}\int^{\infty}_0 d\xi \xi^2 \alpha (i\xi) \mbox{tr}\mathbf{G} (\mathbf{r},\mathbf{r}, i\xi)\nonumber\\
&&-\frac{\mu_0}{3}\omega^2_{10}|\mathbf{d}_{01}|^2\mbox{tr}[\mbox{Re}\mathbf{G}(\mathbf{r},\mathbf{r},\omega_{10})],
\end{eqnarray}
while its chiral components are given by \cite{buh3}
\begin{eqnarray}\label{chiral}
U_{0c}&=&-\frac{\hbar\mu_0}{\pi}\int_0^{\infty}d\xi \xi \Gamma (i\xi) \mbox{tr}[\nabla \times \mathbf{G} (\mathbf{r},\mathbf{r}, i\xi)],\nonumber\\
U_{1c}&=&\frac{\hbar\mu_0}{\pi}\int_0^{\infty}d\xi \xi \Gamma (i\xi) \mbox{tr}[\nabla \times \mathbf{G} (\mathbf{r},\mathbf{r}, i\xi)]\nonumber\\
&&+\frac{2\mu_0\omega_{10}R_{01}}{3}\mbox{tr}[\nabla \times \mbox{Re}[\mathbf{G} (\mathbf{r},\mathbf{r},\omega_{10})]],
\end{eqnarray}
and 
\begin{eqnarray}
\alpha (i\xi) =\frac{2}{3\hbar}\frac{\omega_{10}|\mathbf{d}_{01}|^2}{\omega^2_{10}+\xi^2},\quad  \Gamma (i\xi) =-\frac{2}{3\hbar}\frac{\xi R_{01}}{\omega_{10}^2+\xi^2}.
\end{eqnarray}
Here $\hbar$ is Planck's constant, $\mu_0$ is vacuum permeability, $\mathbf{d}_{01}$ is the electric dipole transition matrix element, and the optical rotatory strength $R_{01}=\mbox{Im} (\mathbf{d}_{01}\cdot \mathbf{m}_{10})$ where $\mathbf{m}_{10}$ is the corresponding magnetic dipole moment matrix element. We have ignored the magnetic component of the potential which is a factor of $(|\mathbf{m}_{10}|/|\mathbf{d}_{01}|)^2\sim \alpha^2_{f}$ smaller than the electric component, and is a factor of $|\mathbf{m}_{10}|/|\mathbf{d}_{01}|\sim \alpha_{f}$ smaller than the chiral component,
with $\alpha_{f}$ being the fine-structure constant. Since the optical rotatory strengths of enantiomers are generally identical in magnitude but opposite in sign, the chiral components of the CP potential (\ref{chiral}) take opposite signs for enantiomers.

Note that, when the average thermal photon number
\begin{eqnarray}
n(\omega_{10})=\frac{1}{e^{\hbar\omega_{10}/(k_{B}T)}-1}
\end{eqnarray}
 is not negligible,  (\ref{electric}) and (\ref{chiral}) can be replaced by the thermal CP potentials which are written as \cite{buh2, sch, sch2, sch3}
 \begin{eqnarray}\label{thermalCP}
 U_{0e}^{\rm therm}&=&\mu_0 k_{B} T\displaystyle\sum_{j=0}^{\infty}\left(1-\frac12\delta_{j0}\right)\xi_{j}'^2  \alpha (i\xi'_{j}) \mbox{tr}\mathbf{G} (\mathbf{r},\mathbf{r},i\xi'_{j})\nonumber\\
 &&+\frac{\mu_0}{3} n (\omega_{10})\omega^2_{10}|\mathbf{d}_{01}|^2\mbox{tr}[\mbox{Re}\mathbf{G}(\mathbf{r},\mathbf{r},\omega_{10})],\nonumber\\
 U_{1e}^{\rm therm}&=&-U_{0e}^{\rm therm}\nonumber\\
 &&-\frac{\mu_0}{3}[n(\omega_{10})+1]\omega_{10}^2|\mathbf{d}_{01}|^2 \mbox{tr}[\mbox{Re} \mathbf{G} (\mathbf{r},\mathbf{r},\omega_{10})],\nonumber\\
 U_{0c}^{\rm therm}&=&-2\mu_0 k_{B}T\displaystyle\sum_{j=0}^{\infty}\left(1-\frac12\delta_{j0}\right)\xi_{j}'  \Gamma (i\xi'_{j}) \nonumber\\
 &&\times \mbox{tr}[\nabla \times \mathbf{G} (\mathbf{r},\mathbf{r},i\xi)]\nonumber\\
 &&-\frac{2\mu_0\omega_{10}R_{01}}{3}n(\omega_{10})\mbox{tr}[\nabla \times \mbox{Re}[\mathbf{G} (\mathbf{r},\mathbf{r},\omega_{10})],\nonumber\\
 U_{1c}^{\rm therm}&=&-U_{0c}^{\rm therm}\nonumber\\
 &&+\frac{2\mu_0\omega_{10}R_{01}}{3}[n(\omega_{10})+1]\mbox{tr}[\nabla \times \mbox{Re}[\mathbf{G} (\mathbf{r},\mathbf{r},\omega_{10})]],\nonumber\\
 \end{eqnarray}
 where $\xi'_{j}=2\pi k_{B} Tj/\hbar$ $(j=0,1,2,\ldots)$ are Matsubara frequencies \cite{matsu}. The thermal CP potentials have contributions proportional to $n(\omega_{10})$ and $n(\omega_{10})+1$ due to the absorption and emission of photons by the molecule respectively. Different from (\ref{electric}) and (\ref{chiral}), we see that a ground-state molecule also exhibits a resonant contribution associated with absorption of thermal photons.

The scattering Green's tensor $\mathbf{G}(\mathbf{r},\mathbf{r},\omega)$ for our cavity setup can be written as \cite{green}
\begin{eqnarray}\label{greentensor}
&&\mathbf{G}(\mathbf{r},\mathbf{r},\omega)=\frac{i}{8\pi^2}\int\frac{d^2 k^{||}}{k^{\perp}}\nonumber\\
&&\times \Biggr[\displaystyle
\sum_{\sigma_1 \sigma_2=s,p}
e^{2ik^{\perp}z}(\mathbf{D}^{-1}\cdot\mathbf{R})^{(\sigma_1,\sigma_2)}\mathbf{e}_{\sigma_1}(k^{\perp})\mathbf{e}_{\sigma_2}(-k^{\perp})\nonumber\\
&&+\displaystyle
\sum_{\sigma_1\sigma_2=s,p}
e^{2ik^{\perp}(a-z)}(\mathbf{R}'\cdot \mathbf{D}^{-1})^{(\sigma_1,\sigma_2)}\mathbf{e}_{\sigma_1}(-k^{\perp})\mathbf{e}_{\sigma_2}(k^{\perp})\Biggr]\nonumber\\
\end{eqnarray}
where the polarization sum runs over $s-$ and $p-$polarizations, $\mathbf{D}=\mathbf{I}-e^{2iak^{\perp}}\mathbf{R}\cdot \mathbf{R}'$, $k^{||}$ and $k^{\perp}$ are the components of the wave vector parallel and perpendicular to the surface of mirrors respectively and $k^{\perp2}=\omega^2/c^2-k^{||2}$, $a$ is the distance between two mirrors (i.e., the width of the cavity). We  discarded terms that are independent of the position of the molecule $z$ ($0<z<a$). Mirror A is located at $z=0$ while mirror B is located at $z=a$. 
$\mathbf{R}=\begin{pmatrix} 
r_{ss} & r_{sp} \\
r_{ps} & r_{pp} 
\end{pmatrix}=\begin{pmatrix} 
-r_{e} & r_{c} \\
-r_{c} & r_{e} 
\end{pmatrix}$ and $\mathbf{R}'=\begin{pmatrix} 
r'_{ss} & r'_{sp} \\
r'_{ps} & r'_{pp} 
\end{pmatrix}=\begin{pmatrix} 
-r_{e} & -r_{c} \\
r_{c} & r_{e} 
\end{pmatrix}$
are the reflection matrices of mirrors A and B, respectively, where the superscript $^{(\sigma_1\sigma_2)}$ represents the element  $(\sigma_1\sigma_2)$ of the matrix. The mirrors are assumed to have opposite chirality. We adopt the convention that a left-handed mirror or a mirror of positive chirality rotates the polarisation of an electromagnetic wave in a counter-clockwise direction upon reflection when travelling alongside the wave, while a right-handed mirror of negative chirality rotates it in a clockwise direction. With this convention, $r_c>0$ implies that mirror A is right-handed or of negative chirality while mirror B is left-handed or of positive chirality. Note that we are neglecting the dependence of the reflection coefficients on frequency and on the in-plane component of the wave vector $\textbf{k}^\parallel$. This is justified as we are mainly interested in the dominant resonant part of the Casimir-Polder potential which is governed by a single transition frequency and travelling-wave contributions at normal incidence. In practice, we will obtain the entries of the reflection matrix from laser reflection experiments on chiral materials currently under development. 

\begin{figure}
{%
\includegraphics[clip,width=1\columnwidth]{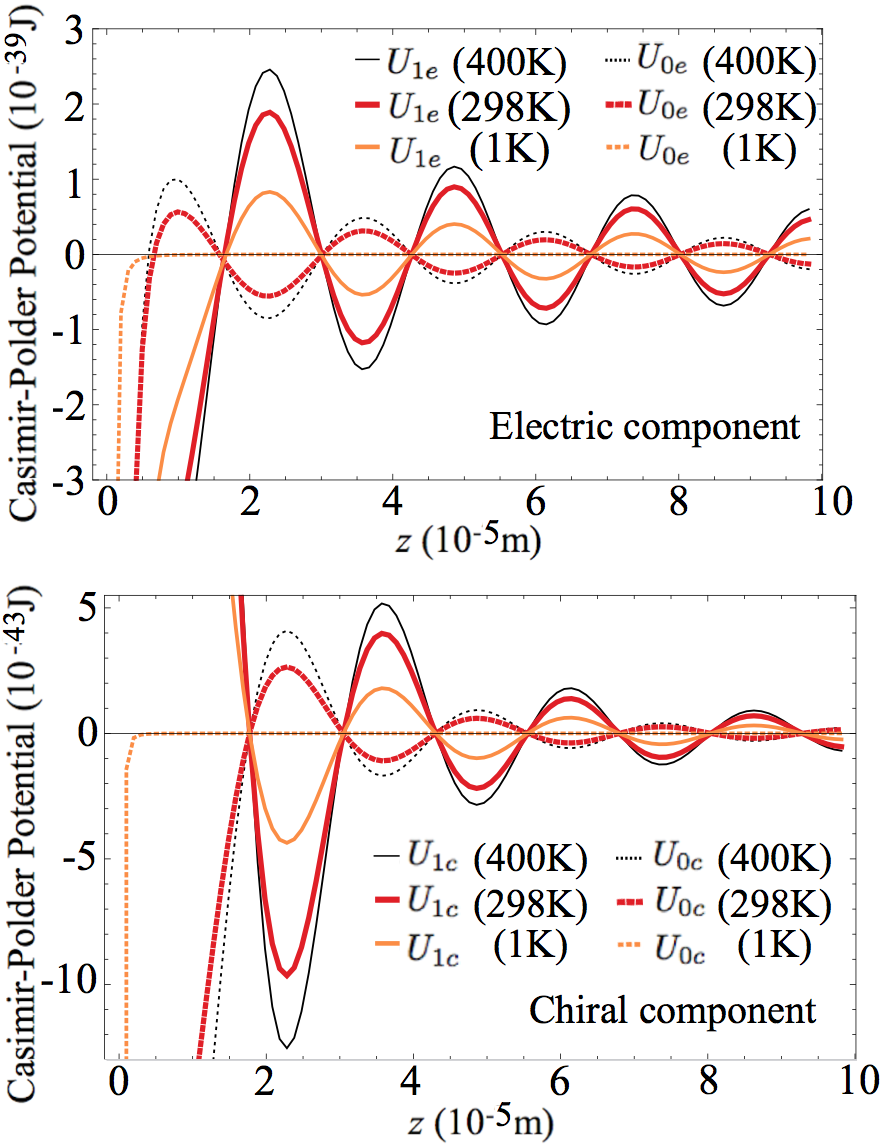}%
}
\caption{(Color online)  The electric and chiral component (experienced by the $(R)$-propylene-oxide molecule with the positive $R_{01}$) of  the CP potential with different temperatures. $r_{e}=0.05,$ $r_{c}=0.8$. The solid lines: the excited-state potential, the dotted lines: the ground-state potential.}
\label{fig6}
\end{figure}
\begin{figure*}
{%
\includegraphics[clip,width=1.5\columnwidth]{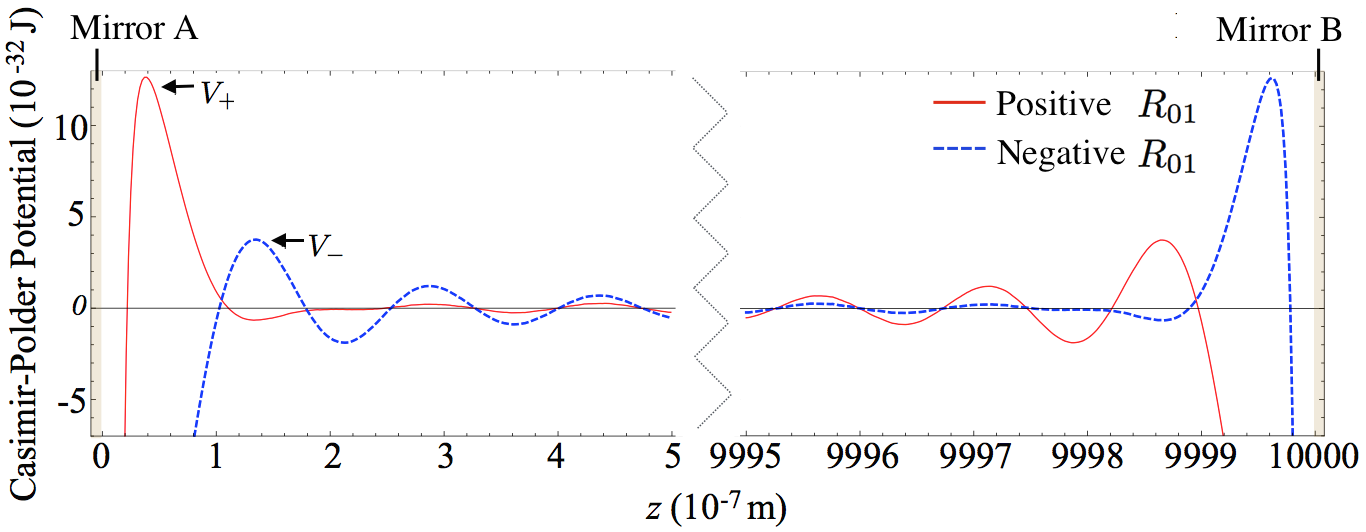}%
}
\caption{(Color online)  The CP potential experienced by the driven 3-MCP molecule in a superposition of half-excited and half-ground state. Solid red line: 3-MCP molecule with the positive $R_{01}$, dotted blue line: 3-MCP molecule with the negative $R_{01}$.}
\label{fig3}
\end{figure*}

In the case of linearly polarized light, the $p$- and $s$- polarized waves  $\mathbf{E}_{s}$, $\mathbf{E}_{p}$ are reflected as $\mathbf{E}_{s}\rightarrow r_{sp} \mathbf{E}_{p}+r_{ss} \mathbf{E}_{s}$ and $\mathbf{E}_{p}\rightarrow r_{pp} \mathbf{E}_{p} + r_{ps} \mathbf{E}_{s}$ by the chiral mirror respectively. Normal mirrors have small or zero values for  $r_{sp}$ and $r_{ps}$.  Perfect chiral mirrors have $|r_{sp}|=|r_{ps}|=1$ and $|r_{ss}|=|r_{pp}|=0$, which exhibits the strongest chiral dependence on the CP potential. For circularly polarized light, normal mirrors in general reverse its polarization upon reflection, but the chiral mirrors conserve the polarization with positive or negative phase shifts.




\subsection{Electric Circular Dichroism}

The  3-MCP molecules 
exhibit electric circular dichroism (ECD) by absorbing left- and right-circularly polarized light differently by electric transitions in the ultraviolet region. We have four isomers of 3-MCP molecules with an equatorial (eq) or axial (ax) methyl group in the $(R)-$ or $(S)-$configurations. For the $S_0\rightarrow S_1$ excitation of the 3-MCP molecule with an equatorial methyl group, we have the electric transition dipole $|\mathbf{d}_{01}|=2.44\times 10^{-31}$Cm, the optical rotatory strength $|R_{01}/c|=8.07\times 10^{-63}\mbox{C}^2\mbox{m}^2$, and the vertical $S_0-S_1$ excitation energy $\hbar\omega_{10} = E_1-E_0=4.24\,$eV ($\omega_{10}=6.44\times 10^{15}\mbox{s}^{-1}$) \cite{3mcp,3mcp2}. Since the optical rotatory strength of the molecule with an equatorial methyl group is slightly larger than that of the molecule with an axial methyl group,  we use the molecule with an equatorial methyl group for a quantitative analysis in this article. Since the enantiomers have the optical rotatory strength of opposite sign and equal in magnitude for every transitions, the chiral component of the CP potential (\ref{chiral}) can be used to separate them. Here  an $(R)$-3-(eq)-MCP molecule and an $(S)$-3-(eq)-MCP molecule have a positive and negative optical rotatory strength $R_{01}$ respectively. Since the ratio $|\mathbf{d}_{01}|^2/|R_{01}/c|=7.38$ is relatively small, the chiral component of the potential is comparable to  its electric component for the molecules. The average thermal photon number is generally very small for the case of ECD (i.e., $n(\omega_{10})\approx 10^{-72}$ for the $S_0\rightarrow S_1$ excitation of the 3-MCP molecule at room temperature $T=298$K), and the temperature dependence of the CP potential is usually negligible for the case of ECD.

The molecule in the excited state has the resonant contributions to the CP potential caused by real photons. When the molecule is far away from the mirrors (i.e., $z, a-z>\lambda_{10}=\frac{2\pi c}{\omega_{10}}$), they  are generally dominant over the non-resonant contributions from virtual photons. For this reason, we introduced a laser field to drive the molecules to excite them. 
In FIG.~\ref{fig2},  the electric component  and  the chiral component of the CP potential experienced by 3-MCP molecule near   mirror A with the parameters above are plotted respectively with the cavity width $a=1\,$mm (Appendix A).  The chiral component in the figure is the one acting on the molecule with the positive $R_{01}$ (i.e., $(R)$-3-(eq)-MCP molecule), and it changes sign for the  molecule with the negative $R_{01}$ (i.e., $(S)$-3-(eq)-MCP molecule). Since   mirror B has opposite chirality from   mirror A, the  molecule sees the chiral component with opposite sign near   mirror B. The magnitude of the electric component and that of the chiral component depend linearly on the reflection coefficient 
$r_{e}$
and
$r_{c}$ respectively. For our purposes, it is desirable to introduce nearly perfect chiral mirrors with $r_{e}\approx 0$ and $r_{c}\approx 1$. For the 3-MCP molecules, the magnitude of the chiral component  is similar to that of the electric component when $r_{c}$ is around 10 times as large as $r_{e}$. 

In \cite{ell}, an enhancement of the the electric component of the CP potential was observed when the cavity width $a$ is equal to a half integer multiple of the molecular transition wavelength, i.e.,  $a=\nu \lambda_{10}/2$, $\nu \in \mathbb{N}$. For the chiral component, it is found that such enhancement occurs when the cavity width $a=\nu  \lambda_{10}/4$ where $\nu$ is even/odd natural number when the two mirrors of the cavity have the same/opposite chirality respectively. In FIG. \ref{fig7}, the electric and chiral component of the CP potential with $r_{e}=0.05$, $r_{c}=0.8$ and the cavity width $a=\nu \lambda_{10}/4$ is plotted. It can be seen that the chiral component is enhanced when $\nu$ is odd, i.e., $\nu=7$, while the electric component increases as $a$ becomes close to a half integer multiple of the molecular transition wavelength. However this enhancement can  only be seen if we have a very small cavity whose width 
contains only a few molecular transition wavelengths, and a strong enhancement only occurs when the reflection coefficients are very close to $1$.

In this article, we choose a realistic setup with $a=1\,$mm where such enhancement can not be observed and rather the CP potential behaves as if it is the sum of two independent  potentials created by two mirrors. Then the potential far away from the mirrors is negligible, and it only becomes significant when the molecule is very close to the mirrors as shown in FIG. \ref{fig2}. Therefore our results presented in this article can be used even for the setup with a single chiral mirror, and the cavity setup here is only used for the purpose of constraining the molecular beam inside the cavity.

\subsection{Vibrational Circular Dichrosim}

In contrast to cases of ECD, we often observe the temperature dependence of the CP potential  caused by thermal photons in the infrared region when we use molecules with vibrational circular dichroism (VCD). In FIG. \ref{fig6}, the temperature dependence of the CP potential experienced by the $(R)$-propylene-oxide molecule with $|\mathbf{d}_{01}|=8.82\times 10^{-32}$Cm, $R_{01}/c=3.89\times 10^{-67}\mbox{C}^2\mbox{m}^2$ (positive optical rotatory strength), and $\omega_{10}=3.8\times 10^{13}\mbox{s}^{-1}$ \cite{pro} is shown as an example. The reflection coefficients are set to be $r_{e}=0.05$, $r_{c}=0.8$.  

Here we have the average thermal photon number $n(\omega_{10})=0.6$ at room temperature and the CP potential increases with temperature. As it can be seen in the figure, the  use of molecules with vibrational circular dichroism has the advantage that the peaks of the CP potential appear far from the mirror due to the long wavelength of photons involved. However it has challenges due to the weakness of the CP potential~(\ref{thermalCP}) which depends on the frequency of photons. Therefore, in the next section, we use the 3-MCP molecules with the stronger CP potential (FIG. \ref{fig2}) in order to analyze our setup to separate enantiomers.

\section{Separation of enantiomers}\label{sec3}

\begin{figure}
{%
\includegraphics[clip,width=1\columnwidth]{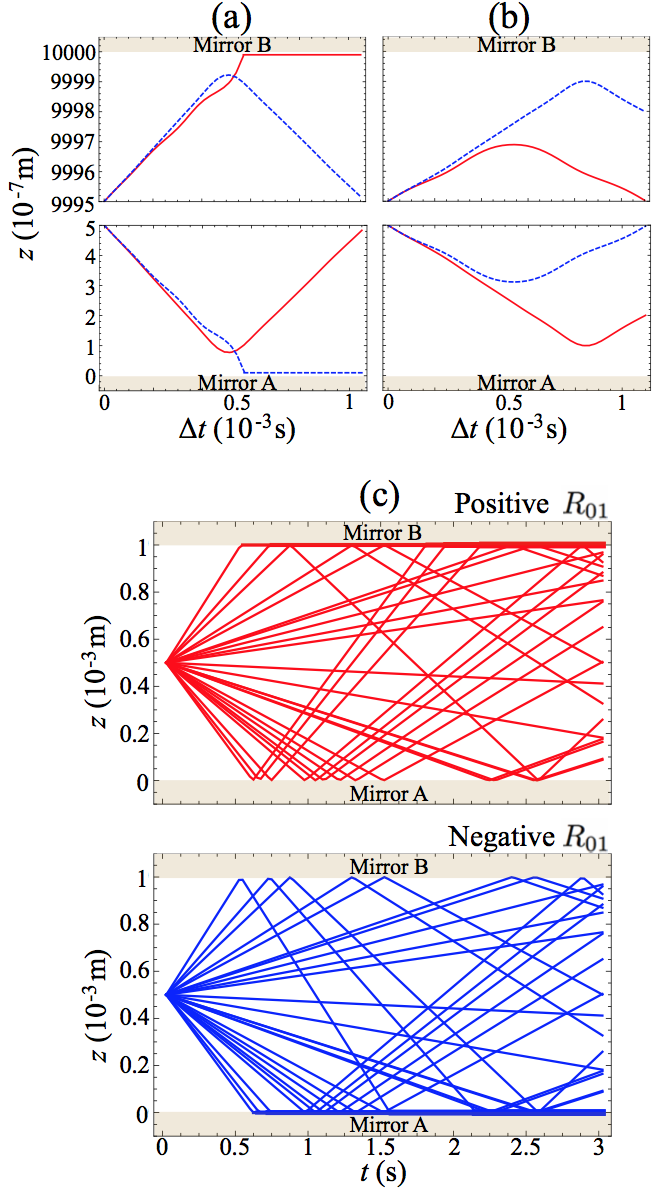}%
}
\caption{(Color online)  Sample trajectories (near the mirrors (a,b)) with different initial velocities. (a,b)  Solid red lines are for the molecules with the positive $R_{01}$ and  dotted blue lines are for those with the negative $R_{01}$.   $v_{z}=+/- 0.9\,$mm/s (above/below) for (a), and  $v_{z}=+/- 0.5\,$mm/s (above/below) for (b). (c) Trajectories of the molecules with the positive $R_{01}$ (above) and those with the negative $R_{01}$(below) with  Gaussian distributed initial velocities with mean $0\,$mm/s and standard deviation $0.4\,$mm/s. (25 trajectories)}
\label{fig4}
\end{figure}
Next, we study trajectories of the 3-MCP molecules  in the cavity. The CP force can be obtained by $F_{CP}=-dU_{CP}/dz$ where $U_{CP}$ is given in (\ref{CPpotential}). We set $r_{c}=0.8$, $r_{e}=0.05$ and the width of the cavity $a=1\,$mm. The mass of one 3-MCP molecule is approximately equal to $1.63\times 10^{-25}$kg. We choose a laser intensity $I=5$W/$\mbox{cm}^2$, a Rabi frequency $\Omega=2|\mathbf{d}_{01}|(2\pi I/c)^{1/2}/\hbar=1.42\times 10^{7}\mbox{s}^{-1}$ \cite{band}, and the detuning $\Delta=\omega_{L}-\omega_{10}=2\pi \times 0.1$MHz. When $\Delta \leq 0.1\Omega$, approximately half of the molecules are in the excited state, and the other half are in the ground state. FIG. \ref{fig3} shows the CP potential experienced by the molecule in such a superposition  state. Near   mirror A, the molecules with the positive $R_{01}$ and those with the negative $R_{01}$ are subject to a high potential barrier $V_{+}$ ($1.26\times 10^{-31}$J) and a low potential barrier $V_{-}$ ($3.76\times 10^{-32}$J) respectively. The differences in the potential barriers seen by these enantiomers come from the chiral component of the CP potential, and the behavior of the potential becomes opposite near  mirror B which has the chirality opposite to that of   mirror A.  $1.26\times 10^{-31}$J and $3.76\times 10^{-32}$J correspond to kinetic energy of the molecule with speed $1.2\,$mm/s and  $0.7\,$mm/s respectively. Therefore, near   mirror A, if the initial speed of the molecule $|v_{z}|$ in the $\hat{z}$-direction (perpendicular to the plane of mirrors) is in between $0.7\,$mm/s and $1.2\,$mm/s, the molecule with the negative $R_{01}$ climbs up the low potential barrier and gets attracted to   mirror A, while the  one with the positive $R_{01}$ can not overcome the high potential barrier and gets repelled back to the opposite side of the cavity (FIG.~\ref{fig4} (a, below)). The opposite behavior of the enantiomers can be observed near   mirror B (FIG.~\ref{fig4} (a, above)). When $|v_{0z}|$ is smaller than $0.7\,$mm/s, both enantiomers can not overcome the potential barrier and get reflected back to the center of the cavity (FIG.~\ref{fig4} (b)). In this way, we aim to collect the molecules with the positive $R_{01}$ and those with the negative $R_{01}$ near   mirror B and   mirror A respectively.

\begin{figure}[b]
{%
\includegraphics[clip,width=0.85\columnwidth]{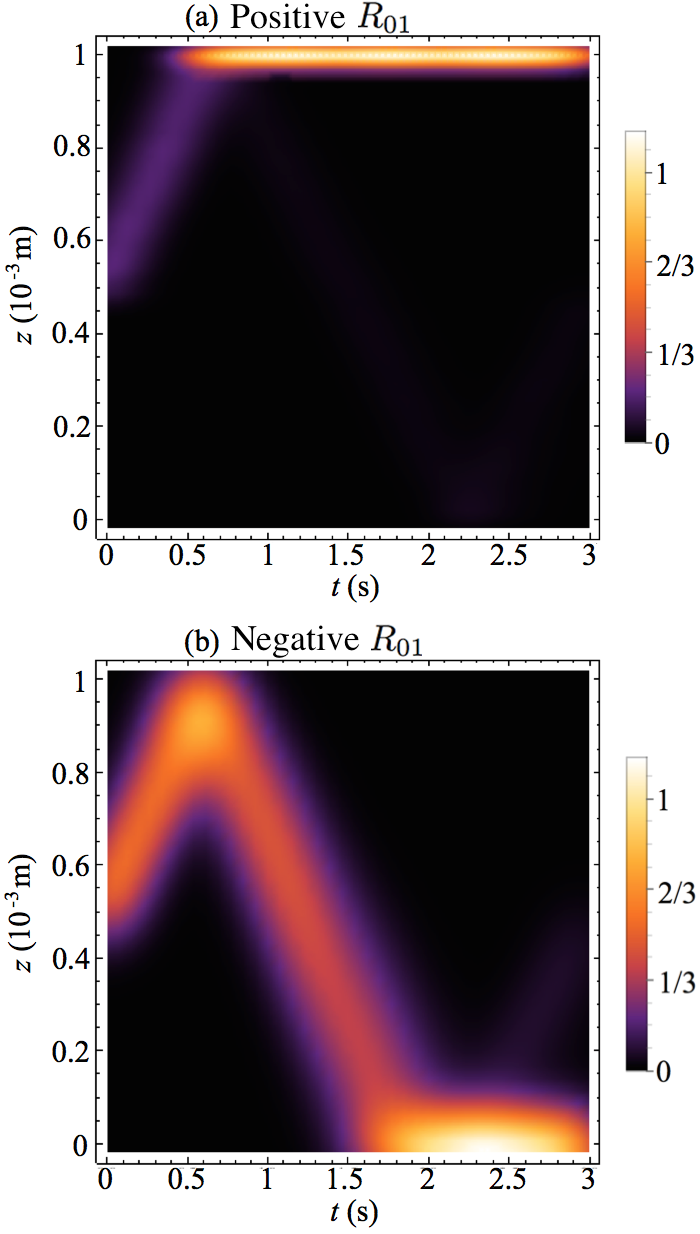}%
}
\caption{(Color online)  Trajectories of (a) the molecules with the positive $R_{01}$ and (b) those with the negative $R_{01}$ whose   initial velocities are Gaussian distributed with mean $0.8\,$mm/s and standard deviation $0.1\,$mm/s. (100 trajectories)   }
\label{fig5}
\end{figure}

We now consider the molecular beam initially injected to the center of the cavity (i.e., initial position of molecule $z_0=0.5\,$mm) with distribution for initial velocities in $\hat{z}$-direction being a Gaussian  with mean $0\,$mm/s and standard deviation $0.4\,$mm/s. The time evolution of  positions of molecules can be obtained numerically (FIG.~\ref{fig4} (c)). As can be seen, $\sim 10\%$ of the molecules with the positive $R_{01}$ and those with the negative $R_{01}$ get collected in the vicinity of   mirror B and of   mirror A respectively after $\sim 1$ second.

A more efficient separation can be realized when the molecules have initial velocities toward one of the mirrors. For example, if the  distribution for initial velocities is a Gaussian with mean $0.8\,$mm/s and a standard deviation $0.1\,$mm/s so that the molecules move toward   mirror B initially, around $90\%$ of the molecules with the positive $R_{01}$ and those with the negative $R_{01}$ get collected near  mirror B and   mirror A respectively after $\sim 1.5$ seconds, with the remaining $10\%$ staying around the center of the cavity (FIG. \ref{fig5}).

\section{Conclusion}\label{sec4}
To summarise, we have presented a realistic proposal for a Stern--Gerlach type  separator for chiral enantiomers based on discriminatory CP forces. We have shown that the enantiomer-selective excited-state CP force due to electric molecular dichroism is most suitable for this purpose. It exhibits pronounced repulsive potential barriers when using chiral mirrors which selectively repel only one of the two enantiomers while attracting the other. By sending a molecular beam through a cavity formed by two mirrors of opposite chirality, each enantiomer gets attracted to one mirror and repelled from the other, leading to an efficient separation. Our analysis shows that our setup can be used for separation of enantiomers, as well as for the detection of the chiral dependence of the CP potential within the reach of current technology. Compared to conventional methods for chiral separations such as chromatography, our method proposed here is more universal since the same chiral mirror can be used for chiral separation of a wide range of molecules.

Two alternatives might be considered in order to potentially improve our scheme. First, while we use an exciting laser parallel to the plates, one could also consider an evanescent laser field emerging from the chiral mirrors themselves. The resulting light force could possibly be designed to enhance chiral effects. Secondly, the use of curved mirrors would lead to a stronger mode confinement and associated enhancement of the CP potential. This could be particularly useful for molecules with vibrational dichroism, where the potential is weak to start with, but cavity enhancement can be made particularly strong as the respective transitions are in the microwave regime.

\begin{acknowledgments}

We would like to thank P. Barcellona, R. Bennett, S. Fuchs and A. Salam for discussions. This work was supported by the German Research
Council (grants BU1803/3-1 and  GRK 2079/1), a National Science and Engineering Research Discovery Grant in Canada, funds from Canada Foundation for Innovation for the Centre for Research on Ultra-Cold Systems (CRUCS) and Chirality Research on Origins and Separation (CHIROS) at UBC. F. S. thanks DAAD Research Grants for support. S.Y.B is grateful for support by the Freiburg Institute of Advanced Studies.

\end{acknowledgments}

\appendix

\section{Derivation of the Casimir-Polder Potential}
Here we derive the electric and the chiral components of the Casimir-Polder (CP) potential experienced by a molecule in a planar cavity.

Taking the trace of the Green's function is equivalent to taking the dot product between the dyads where
\begin{eqnarray}
\mathbf{e}_{s}\cdot \mathbf{e}_{p} (-k^{\perp})=\mathbf{e}_{p}(k^{\perp})\cdot \mathbf{e}_{s}=0
\end{eqnarray}
and
\begin{eqnarray}
\mathbf{e}_{s}\cdot \mathbf{e}_{s}=1, \quad \mathbf{e}_{p}(k^{\perp})\cdot \mathbf{e}_{p}(-k^{\perp})=-1+\frac{2k^{||2}c^2}{\omega^2}.
\end{eqnarray}

Since $\int d^2 k^{||}=2\pi \int_0^{\infty} k^{||}dk^{||}$, we have
\begin{eqnarray}
&&\mbox{tr}\mathbf{G}(\mathbf{r},\mathbf{r},\omega)=\frac{i}{4\pi}\int\frac{d k^{||}k^{||}}{k^{\perp}}\nonumber\\
&&\times \Biggr[e^{2ik^{\perp}z}\left((\mathbf{D}^{-1}\cdot\mathbf{R})^{(s,s)}-\left(1-\frac{2k^{||2}c^2}{\omega^2}\right)(\mathbf{D}^{-1}\cdot\mathbf{R})^{(p,p)}\right)\nonumber\\
&&+e^{2ik^{\perp}(a-z)}\nonumber\\
&&\times\left((\mathbf{R}'\cdot \mathbf{D}^{-1})^{(s,s)}-\left(1-\frac{2k^{||2}c^2}{\omega^2}\right)(\mathbf{R}'\cdot \mathbf{D}^{-1})^{(p,p)}\right)\Biggr]
\end{eqnarray}
where we discarded the terms that are independent of the position of the molecule $z$ since we are eventually interested in the CP force given by $F_{CP}=-d U_{CP}/d z$.

When we introduce $\omega=i\xi$ and $k^{\perp}=i\kappa^{\perp}$ ($\kappa^{\perp}=\sqrt{\xi^2/c^2+k^{||2}}$), we have $\int_0^{\infty} dk^{||} k^{||}/\kappa^{\perp}=\int_{\xi/c}^{\infty}d\kappa^{\perp}$ and
\begin{eqnarray}
&&\mbox{tr}\mathbf{G}(\mathbf{r},\mathbf{r},i\xi)=\frac{1}{4\pi}\int_{\xi/c}^{\infty} d\kappa^{\perp}\nonumber\\
&&\times \Biggr[e^{-2\kappa^{\perp}z}\left((\mathbf{D}^{-1}\cdot\mathbf{R})^{(s,s)}+\left(1-\frac{2\kappa^{\perp 2}c^2}{\xi^2}\right)(\mathbf{D}^{-1}\cdot\mathbf{R})^{(p,p)}\right)\nonumber\\
&&+e^{-2\kappa^{\perp}(a-z)}\nonumber\\
&&\times\left((\mathbf{R}'\cdot \mathbf{D}^{-1})^{(s,s)}+\left(1-\frac{2\kappa^{\perp2}c^2}{\xi^2}\right)(\mathbf{R}'\cdot \mathbf{D}^{-1})^{(p,p)}\right)\Biggr].
\end{eqnarray}

Therefore,
\begin{eqnarray}
&&U_{0e}=\frac{\hbar\mu_0}{8\pi^2}\int_0^{\infty}d\xi \xi^2 \alpha (i\xi)\int_{\xi/c}^{\infty} d\kappa^{\perp}\nonumber\\
&&\times \Biggr[e^{-2\kappa^{\perp}z}\left((\mathbf{D}^{-1}\cdot\mathbf{R})^{(s,s)}+\left(1-\frac{2\kappa^{\perp 2}c^2}{\xi^2}\right)(\mathbf{D}^{-1}\cdot\mathbf{R})^{(p,p)}\right)\nonumber\\
&&+e^{-2\kappa^{\perp}(a-z)}\nonumber\\
&&\times\left((\mathbf{R}'\cdot \mathbf{D}^{-1})^{(s,s)}+\left(1-\frac{2\kappa^{\perp2}c^2}{\xi^2}\right)(\mathbf{R}'\cdot \mathbf{D}^{-1})^{(p,p)}\right)\Biggr],
\end{eqnarray}
and
\begin{eqnarray}
&&U_{1e}=-U_{0e}\nonumber\\
&&-\frac{\mu_0}{12\pi}\omega^2_{10}|\mathbf{d}_{01}|^2\mbox{Re}\Biggr\{i\int_0^{\infty}\frac{d k^{||}k^{||}}{k^{\perp}}\nonumber\\
&&\times \Biggr[e^{2ik^{\perp}z}\left((\mathbf{D}^{-1}\cdot\mathbf{R})^{(s,s)}-\left(1-\frac{2k^{||2}c^2}{\omega^2_{10}}\right)(\mathbf{D}^{-1}\cdot\mathbf{R})^{(p,p)}\right)\nonumber\\
&&+e^{2ik^{\perp}(a-z)}\nonumber\\
&&\times\left((\mathbf{R}'\cdot \mathbf{D}^{-1})^{(s,s)}-\left(1-\frac{2k^{||2}c^2}{\omega^2_{10}}\right)(\mathbf{R}'\cdot \mathbf{D}^{-1})^{(p,p)}\right)\Biggr]\Biggr\}.\nonumber\\
\end{eqnarray}

For the chiral components of the potential, we use $\mathbf{a}\times \mathbf{bc}=(\mathbf{a}\times\mathbf{b})\mathbf{c}$ \cite{chen}, then
\begin{eqnarray}
\nabla \times \mathbf{e}_{s}=-i\frac{\omega}{c} \mathbf{e}_{p} (k^{\perp}), \quad \nabla \times \mathbf{e}_{p}(k^{\perp})=i\frac{\omega}{c}\mathbf{e}_{s}.
\end{eqnarray}

 Therefore,
\begin{eqnarray}
&&\mbox{tr}[\nabla\times \mathbf{G}(\mathbf{r},\mathbf{r},\omega)]=\frac{\omega}{4\pi c}\int\frac{d k^{||}k^{||}}{k^{\perp}}\nonumber\\
&&\times \Biggr[e^{2ik^{\perp}z}\left((\mathbf{D}^{-1}\cdot\mathbf{R})^{(s,p)}\left(\frac{2k^{||2}c^2}{\omega^2}-1\right)-(\mathbf{D}^{-1}\cdot\mathbf{R})^{(p,s)}\right)\nonumber\\
&&+e^{2ik^{\perp}(a-z)}\Bigr((\mathbf{R}'\cdot \mathbf{D}^{-1})^{(s,p)}\left(\frac{2k^{||2}c^2}{\omega^2}-1\right)\nonumber\\
&&\qquad\qquad\qquad\qquad\qquad\qquad\qquad-(\mathbf{R}'\cdot \mathbf{D}^{-1})^{(p,s)}\Bigr)\Biggr],
\end{eqnarray}
and
\begin{eqnarray}
&&\mbox{tr}[\nabla\times \mathbf{G}(\mathbf{r},\mathbf{r},i\xi)]=-\frac{\xi}{4\pi c}\int^{\infty}_{\xi /c}d\kappa^{\perp}\nonumber\\
&&\times \Biggr[e^{-2\kappa^{\perp}z}\left((\mathbf{D}^{-1}\cdot\mathbf{R})^{(s,p)}\left(\frac{2\kappa^{\perp 2}c^2}{\xi^2}-1\right)+(\mathbf{D}^{-1}\cdot\mathbf{R})^{(p,s)}\right)\nonumber\\
&&+e^{-2\kappa^{\perp}(a-z)}\Bigr((\mathbf{R}'\cdot \mathbf{D}^{-1})^{(s,p)}\left(\frac{2\kappa^{\perp 2}c^2-1}{\xi^2}\right)\nonumber\\
&&\qquad\qquad\qquad\qquad\qquad\qquad\qquad+(\mathbf{R}'\cdot \mathbf{D}^{-1})^{(p,s)}\Bigr)\Biggr].
\end{eqnarray}

Finally we obtain
\begin{eqnarray}
&&U_{0c}=\frac{\hbar\mu_0}{4\pi^2c}\int_0^{\infty}d\xi \xi^2 \Gamma (i\xi)\int^{\infty}_{\xi /c}d\kappa^{\perp}\nonumber\\
&&\times \Biggr[e^{-2\kappa^{\perp}z}\left((\mathbf{D}^{-1}\cdot\mathbf{R})^{(s,p)}\left(\frac{2\kappa^{\perp 2}c^2}{\xi^2}-1\right)+(\mathbf{D}^{-1}\cdot\mathbf{R})^{(p,s)}\right)\nonumber\\
&&+e^{-2\kappa^{\perp}(a-z)}\Bigr((\mathbf{R}'\cdot \mathbf{D}^{-1})^{(s,p)}\left(\frac{2\kappa^{\perp 2}c^2-1}{\xi^2}\right)\nonumber\\
&&\qquad\qquad\qquad\qquad\qquad\qquad\qquad+(\mathbf{R}'\cdot \mathbf{D}^{-1})^{(p,s)}\Bigr)\Biggr],
\end{eqnarray}
and
\begin{eqnarray}
&& U_{1c}=-U_{0c}\nonumber\\
&&+\frac{\mu_0\omega_{10}^2R_{01}}{6\pi c}\mbox{Re}\Biggr\{\int\frac{d k^{||}k^{||}}{k^{\perp}}\nonumber\\
&&\times \Biggr[e^{2ik^{\perp}z}\left((\mathbf{D}^{-1}\cdot\mathbf{R})^{(s,p)}\left(\frac{2k^{||2}c^2}{\omega^2_{10}}-1\right)-(\mathbf{D}^{-1}\cdot\mathbf{R})^{(p,s)}\right)\nonumber\\
&&+e^{2ik^{\perp}(a-z)}\Bigr((\mathbf{R}'\cdot \mathbf{D}^{-1})^{(s,p)}\left(\frac{2k^{||2}c^2}{\omega^2_{10}}-1\right)\nonumber\\
&&\qquad\qquad\qquad\qquad\qquad\qquad\quad-(\mathbf{R}'\cdot \mathbf{D}^{-1})^{(p,s)}\Bigr)\Biggr]\Biggr\}.
\end{eqnarray}

\end{document}